\documentstyle[psfig,aps,prl,amsfonts,amssymb,epsf]{revtex}

\begin{document}
\title{Bures/Statistical Distinguishability 
 Probabilities of Triseparable and Biseparable
Eggeling-Werner States}
\author{Paul B. Slater}
\address{ISBER, University of
California, Santa Barbara, CA 93106-2150\\
e-mail: slater@itp.ucsb.edu,
FAX: (805) 893-7995}

\date{\today}

\draft
\maketitle
\vskip -0.1cm

\begin{abstract}
In a number of previous studies, we have investigated the use of the volume
element of the Bures (minimal monotone) 
metric --- identically, one-fourth of the {\it statistical distinguishability} 
(SD) metric --- as a natural measure over the $n^2-1$-dimensional convex set of $n \times n$ density matrices. This has led us for the cases
$n=4$ and 6 
to estimates of the prior (Bures/SD) probabilities that 
 qubit-qubit and qubit-qutrit pairs are {\it separable}. Here, we extend this work
from such {\it bipartite} systems to the {\it tripartite} 
``laboratory'' 
quantum systems
possessing $U \otimes U \otimes U$ symmetry recently constructed by
Eggeling and Werner (Phys. Rev. A 63 [2001], 042324).
We derive the associated SD metric tensors for the three-qubit and
three-qutrit cases, and then obtain estimates of the various 
related Bures/SD
probabilities using Monte Carlo methods.
\end{abstract}

\vspace{.1cm}

\hspace{1cm} {\bf{Key words:}} Qubits, qutrits, density matrix, Bures
metric, statistical distinguishability metric, entanglement, Eggeling-Werner
tripartite state,
Monte Carlo simulation

\vspace{.15cm}

\section{Introduction}
In a number of previous studies \cite{paul1,paul2,paul3,paul4} --- developing 
upon an initial idea of
\.Zyzckowski, Horodecki, Sanpera and Lewenstein \cite{zycz1} --- we have 
investigated
the use of the volume element of the Bures metric \cite{hub1,hub2,braun} 
as a natural measure
on the quantum states. In particular, we have been interested
in its use to evaluate the {\it a priori} probability that a 
(bipartite) state is
classical in nature (separable) or, equivalently, the 
complementary probability
 that it is quantum (nonseparable/entangled).
The Bures metric serves as the {\it minimal} monotone metric \cite{petzsudar}.
The probabilities of separability its volume element provide, appear to be
{\it larger} than for any other member of the nondenumerable class of monotone metrics. (Classically, there is a {\it unique} monotone metric --- the {\it Fisher 
information} metric \cite{kass,frieden}.)
 In this sense, it also constitutes an {\it upper}
 bound on other
(arguably plausible) prior 
probabilities of separability --- while the {\it maximal} monotone metric
\cite{yl,helstrom} would provide  a {\it lower} bound.
(The Bures metric is associated with the {\it symmetric} logarithmic
derivative and the maximal monotone metric with the {\it right}
logarithmic derivative \cite{holevo}.)

In \cite{paul2}, we found
for certain low-dimensional scenarios involving two qubits, {\it analytically
exact} Bures probabilities of separability. This further led us 
in \cite{paul3} to use numerical methods (quasi-Monte Carlo integration), employing a recently-developed 
parameterization of the $4 \times 4$ density matrices 
 based on $SU(4)$ Euler angles
\cite{tbs}, 
to investigate the conjecture that the Bures probability of separability of two {\it arbitrarily}
coupled qubits --- occupying a 15-dimensional convex set --- might also be
exact. In \cite{silver}, we found numerical support for the conjecture that the Bures probability of separability of two qubits is $\frac{1680 \sigma_{Ag}}{\pi^8} \approx 0.0733389$, where
$\sigma_{Ag}$ is the silver mean, $\sqrt{2} -1$.
(Our {\it primary}
 conjecture there was that the volume --- as measured in terms
of (four times) the Bures metric, that is, the statistical distinguishability
(SD) metric \cite{braun} --- of the separable two-qubit states is
${\sigma_{Ag} \over 3}$. Certain other monotone metrics of interest also
appear to give volumes of the separable states that are simple multiples
of $\sigma_{Ag}$, such as $10 \sigma_{Ag}$ for 
the Kubo-Mori metric \cite{silver}. These analyses, however, cast some doubt
as to the naturalness of our earlier conjectures in \cite{paul4} in regard to
the Bures probability of separability of a qubit and {\it qutrit}.)

The development by Eggeling and Werner 
\cite{egg} of new (``laboratory'')  systems of {\it tripartite}
 states --- composed of three subsystems of equal but arbitrary
finite Hilbert space dimension ($d$) ---  provides us with an opportunity
 to take this line of analysis to a {\it finer} level of description
than previously, that is, by attaching Bures volumes to the sets of
biseparable, triseparable states, as well as those having positive partial
transpose with respect to the biseparable partition (cf. \cite{acin}).
These states were devised by Eggeling and Werner to obtain a {\it dimension-independent} characterization of the separability properties of symmetric states.
They possess an explicit parameterization as linear combinations of 
permutation operators, which is helpful in explicit computations.
Also, there is a ``twirl'' operation, which brings an arbitrary tripartite
state to this special subset.

It has proved convenient in previous analyses to speak in terms of the
SD metric rather than the Bures one, and we will follow such a course
here.  All the probabilities given below will, of course,
 be the same using 
either the Bures
or SD metric.
As a point of reference, for the {\it bipartite} (one-dimensional laboratory) 
Werner states \cite{werner} composed of two 
qubits, the
Bures/SD probability of separability was found to be {\it exactly}
${1 \over 4}$ \cite[sec.II.A.7]{paul2}.

\section{Analyses}

Five parameters ($r_{-},r_{+},r_{1},r_{2},r_{3}$) 
characterize the Eggeling-Werner (EW) states, in general, but
only four in the three-qubit case (where the parameter $r_{-}$ 
degenerates to identically
zero). (Also, it is convenient to employ a ``dummy'' bound variable, 
$r_{0} = 1- r_{-}-r_{+}$, as in \cite{egg}, as well as utilize the 
transformation to spherical coordinates, 
$r_{1} = R \cos{\theta}, r_{2}= R \sin{\theta}
\cos{\phi}, r_{3} =R \sin{\theta} \sin{\phi}$.)
For an EW-density matrix, the parameters obey the relations
\cite[eq. (6)]{egg},
\begin{equation} \label{e4}
r_{+},r_{-},r_{0} \geq 0, \quad r_{+}+r_{-}+r_{0} = 1,\quad
r_{1}^2 +r_{2}^2 +r_{3}^2 \leq r_{0}^2.
\end{equation}
The eigenvalues of an EW-density matrix are of the form,
\begin{equation}
{r_{+} \over \nu_{+}} \quad (\mbox{multiplicity} \quad \nu_{+}),\quad
{r_{-} \over \nu_{-}} \quad (\mbox{m}. \quad \nu_{-}), \quad
{1 \over 2 \nu_{0}} (r_{0}  \pm \sqrt{r_{1}^2 +r_{2}^2 +r_{3}^3})
\quad (\mbox{m}. \quad \nu_{0}),
\end{equation}
where
\begin{equation}
\nu_{+}= {d^3+3 d^2 + 2 d \over 6},\quad \nu_{-} = {d^3 -3 d^2 + 2 d \over
6},\quad \nu_{0} = {d^3 -d \over 3}.
\end{equation} 
(The multiplicity $\nu_{+}$ is equal to the dimension of the corresponding
Bose subspace divided by the dimension --- that is, 1 -- of the
respective [trivial] irrep, while $\nu_{-}$ similarly corresponds 
to the Fermi
subspace and $\nu_{0}$ to the para-subspace.)
So, for $d=2$ there are, in general, three distinct eigenvalues, and
for $d>2$, four distinct eigenvalues.

Since we were able rather 
readily to determine, in addition, the corresponding eigenvectors
of the EW states ($\rho$) we examined, 
we could  {\it directly} implement the basic formula for the
Bures metric,
\begin{equation} \label{e1}
d_{B}(\rho,\rho+\mbox{d} \rho)^2 = 
{1 \over 2} \sum_{\alpha,\beta} {|<\alpha |\mbox{d} \rho | \beta >|^2 \over 
\lambda_{\alpha} + \lambda_{\beta}},
\end{equation}
(where $|\alpha>, \alpha = 1,\ldots,n$, are eigenvectors of $\rho$ with
eigenvalues $\lambda_{\alpha}$)
without having to rely upon any of the interesting 
(indirect) 
methods Dittmann developed 
\cite{ditt1} --- and we have applied elsewhere \cite{paul2,paul4,paul5} --- in 
order to avoid the (often problematical) computation of eigenvalues
and eigenvectors of $\rho$.
(For the SD metric we replace the coefficient ${1 \over 2}$ in (\ref{e1})
with 2.) We also observe the relationship 
\begin{equation}
d_{B}(\rho_{1},\rho_{2})^2  =
2 - 2 \sqrt{F(\rho_{1},\rho_{2})},
\end{equation}
where $F(\rho_{1},\rho_{2})$ is the 
frequently-used {\it Bures fidelity} between density 
matrices $\rho_{1}$ and $\rho_{2}$. \cite{hub1,jozsa,ungar}.)

We first examined the EW-states in the qubit ($d=2$) and qutrit ($d=3$) cases.
It turned out that the Bures metric tensor in the qubit (four-parameter
[$r_{-}=0$])
case is {\it identical} to the corresponding $4 \times 4$ submatrix of 
the Bures metric tensor in the qutrit (five-parameter) case.
Also, numerical evidence we have adduced strongly indicates that the
$5 \times 5$ 
metric tensor (and hence the volume element) in the case of three
{\it four}-level ($d=4$) quantum systems is {\it identical} to that in the
three-qutrit case.
Presumably, this holds true for tripartite
EW-states with still higher-dimensional
subsystems $(d>4$).
(In fact, T. Eggeling has been able to independently 
confirm this proposition, as well as the 
formulas (\ref{gfirst})-(\ref{e3}) given immediately below by 
deriving a {\it general} formula --- in terms 
of the subsystem dimension $d$ --- for the Bures metric tensor of
the EW tripartite states, on the basis of {\it representation
theory} (App. I). He has also been able to show that the scalar curvature of
the Bures metric is equal to $20+18/r_{0}$, and thus diverges for $r_{0}=0$
 (cf. \cite{jochensc}).) ``Surprisingly, it turns out that the separability
sets we investigate are also independent of dimension'' \cite{egg}.

In terms of the spherical coordinates, the SD metric tensor
elements ($g_{ij}$) which are not identically zero are the simple functions,
\begin{equation} \label{gfirst}
g_{r_{-}  r_{-}} = 
{1 \over 2} \Big({2 \over r_{-}} + {1 \over r_{0} +R}
-{1 \over -r_{0} +R }\Big),
\end{equation}
\begin{equation}
g_{r_{-} r_{+}} ={r_{0} \over (-r_{0}-R)(-r_{0} +R)},
\end{equation}
\begin{equation}
g_{r_{-} R} = -{R \over (r_{0}+R) (-r_{0}+R)},
\end{equation}
\begin{equation}
g_{r_{+} r_{+}} = {1 \over 2} \Big( {2 \over r_{+}} +{1 \over r_{0} +R}
-{1 \over -r_{0}+R} \Big),
\end{equation}
\begin{equation}
g_{r_{+} R} = - {R \over (r_{0}+R) (-r_{0}+R)},
\end{equation}
\begin{equation}
g_{R R} = {r_{0} \over (-r_{0} -R) (-r_{0} +R)},
\end{equation}
\begin{equation}
g_{\theta \theta} = {R^2 \over r_{0}},
\end{equation}
\begin{equation} \label{glast}
g_{\phi \phi} = {R^2 \sin^{2}{\theta} \over r_{0}}.
\end{equation}

The SD volume element ($|g_{ij}|^{1/2}$)
in the {\it qubit} case (first, having set $r_{-} =0$ in the above
and deleted
the row and column of the $5 \times 5$ metric tensor corresponding to 
$r_{-}$ before computing  $|g_{ij}|^{1 \over 2}$) is
(reverting to the original EW coordinates),
\begin{equation} \label{e2}
|g_{ij}|^{1/2}_{qubit} = {1 \over r_{0}  
\sqrt{r_{+} \Big(-r_{1}^2 -r_{2}^2 -(r_{0} +r_{3}) (-r_{0} +r_{3})\Big)}},
\end{equation}
and, quite similarly, in the qutrit case,
\begin{equation} \label{e3}
g_{ij}|^{1/2}_{qutrit} = 
{1 \over r_{0} 
\sqrt{r_{-} r_{+} \Big(-r_{1}^2 -r_{2}^2 - (r_{0} +r_{3} )
(-r_{0} +r_{3}) \Big)}}.
\end{equation}
To normalize (\ref{e2}) to a probability distibution over the 
EW qubit-qubit-qubit states, one must 
divide it
by $2 \pi^{2}/3$, while to normalize (\ref{e3}) over the
EW qutrit-qutrit-qutrit states, one divides it by
$\pi^{3} /2$.

Eggeling and Werner also gave explicit parameter ranges, so that $\rho$
: (1)  is biseparable with respect to the partition $1|23$; (2) is 
triseparable; 
and (3) has a partial transpose that is positive with respect to the first
tensor factor.
We have used these explicit ranges in numerical (Monte Carlo) simulations
to estimate the relative probabilities that an EW-state has any of these
three special properties. In the qubit ($r_{-}=0$) 
case, however, biseparability simply
implies that the corresponding partial transpose is positive \cite{egg}.

\subsection{Probabilities for qutrit case}
For the three-{\it qutrit} case, based on some 131 million 
randomly generated points, 
an estimate of .0963689  
 was gotten for the SD/Bures  probability of a positive
partial transpose, .0694443  for the SD/Bures probability of biseparability,
and .0165952  for triseparability.
The sample was compiled in {\it five} roughly equal subsample. (Actually, for 
each subsample, one {\it billion} random uniformly distributed 
points --- $r_{-}$ and $r_{+}$ being drawn 
from [0,1] and $r_{1},r_{2},r_{3}$ from
[-1,1] --- were tested to ascertain if they
fullfilled the requirements (\ref{e4}) for an EW-state. 
Approximately 2.6 percent
 did in each
subsample.) In Table I, we also 
show the results for each subsample, so one 
might
gauge the stability of the overall results. (The probabilities were computed
by taking the {\it ratio} of the SD volume assigned to the particular subset of
states in question to the SD volume assigned to all those approximately 131 million states meeting
the requirements for a general EW-state.) The standard
deviations over the five samples are also given --- assuming the nearly
equal sample sizes are, in fact, strictly equal.
(Since the true probabilities are unknown, but only estimated, 
we use the {\it bias-adjusted}
standard deviation, ${\sqrt{\Sigma_{i=1}^{5} (x_{i}-\bar{x})^2 \over n}}$, 
with $n = 4$
rather than $n=5$.)

Based on the results in Table I, it would perhaps 
appear to be somewhat of an
overstatement ``that the two sets [biseparable states and states with
positive partial transposes] come to be remarkably close (see. Fig. 11)''
\cite[p. 10]{egg}.
\begin{table}
\begin{tabular}{r|r|r|r|}
\hline
sample size & Bures/SD prob. of positive  part. transp. & Bures/SD prob. of biseparable  & Bures/SD prob. of triseparable \\
\hline
\hline
26,176,260 & .0960822 & .0692970 & .0141430 \\
\hline
26,177,694 &  .0962687  & .0691823 & .0142139 \\
\hline
26,179,742 & .0964395 & .0696499 & .0143551  \\
\hline
26,183,533 & .0963464 & .0693287 & .0141516  \\
\hline
26,173,004 & .0967076 & .0697630 & .0144027 \\
\hline
\hline
130,890,233 & .0963689  &  .0694443 & .0142526  \\
\hline
std. dev. & .00023 & .000249 & .000119 \\
\end{tabular}
\caption{Estimated Bures/SD probabilities, based on Monte Carlo simulations, 
 that a three-qutrit EW-state:
(1) has a positive partial transpose with respect to the partition
$1|23$; (2) is biseparable with respect to the partition; (3) 
is triseparable; and (4) is biseparable with respect to the 
partition {\it given} that it is 
triseparable. For each of the five subsamples, one billion points were 
initially generated, from which those corresponding to EW-states were 
selected.}
\end{table}

\.Zyczkowski, Horodecki, Sanpera and Lewenstein, in the conclusion of
their pioneering paper \cite{zycz1}, had asked the question:
``Has the set of separable states really a volume strictly smaller than
the volume of the set of states with a positive partial transpose?''
So, their question is answered affirmatively in the context here.

In our initial attempts to  calculate the entries of Table I, however, 
we had been quite perplexed to find
that roughly 14 percent of the probability mass assigned to the triseparable
states appeared {\it not} to  meet the EW-criteria for biseparability.
(A randomly generated 
example of one such state had parameters $r_{+} = .27, r_{-}=.1,r_{1} =
.589304, r_{2} = .08100014,r_{3} = -.138433$.)
After correspondence with T. Eggeling regarding this 
clearly paradoxical situation,
he concluded that it was necessary to incorporate --- which we proceeded 
to do, as reflected in Table I exhibited here --- the 
{\it additional} constraint,
\begin{equation} \label{new}
r_{1}^2 +r_{2}^2 +r_{3}^2 \leq 4 (r_{+}-r_{-})^2,
\end{equation}
into the set of three already given in \cite[sec. III]{egg}.
The problem encountered, Eggeling indicated, 
 stemmed from the fact that the third
of the published constraints delimiting the domain of triseparability 
is of the {\it third}-degree. 
The added constraint 
(\ref{new}) is
necessary to ensure convexity. The plots in \cite{egg} other than Fig.~5 are
themselves unaffected, he noted, 
since an appropriate numerical cutoff that took the ``right'' root
of the third-degree polynomial was utilized in their generation.
Fig.~1  
is the same as Fig.~5 in \cite{egg}, except that in addition to the 
central heart-shaped region of triseparable states displayed there, now there are also 
shown the three peripheral regions that the new fourth constraint (\ref{new})
formally excludes.
\begin{figure}
\centerline{\psfig{figure=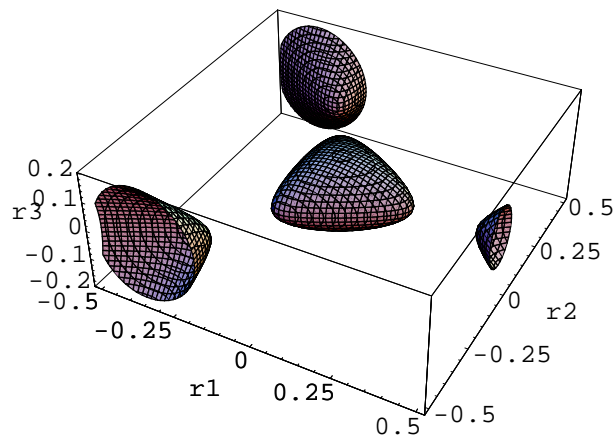}}
\caption{Plot for the section $r_{+} =0.27$ and $r_{-}=0.1$ of the 
(central) heart-shaped set of triseparable states, along with the three
peripheral regions that are excluded by the {\it new} fourth 
triseparability constraint
(\ref{new}). {\it Without} these three 
 peripheral regions, the figure is identical to 
Fig. 5 in [9]}
\end{figure}

\subsection{Probabilities for qubit case} 

Based on some 654 million sampled
points, we obtained in the three-{\it qubit}
 case, an estimate
of .216769 that an EW-state is 
biseparable (with respect to the partition $1|23$)
and .0630532 that it is triseparable.
(As observed earlier, in this specific
case, biseparability implies a positive partial transpose \cite{egg}.)
Table II is the qubit analogue of Table I.
(Again, one billion random uniformly distributed points were examined in each 
randomly generated 
subsample to ascertain if they parameterized an EW-three-qubit density
matrix. Approximately, 13.1 percent did in each subsample.)
\begin{table}
\begin{tabular}{r|r|r|}
\hline
sample size & Bures/SD prob. of biseparable state  & Bures/SD prob. of triseparable state\\
\hline
\hline
130,866,541  & .216612 & .0631270  \\
\hline
 130,860,088 & .216746  & .0630338 \\
\hline
  130,871,093 & .216888 & .0631249 \\
\hline
130,852,647  & .216821 & .0630014  \\
\hline
 130,867,972 &  .216779  &  .0629789 \\
\hline
\hline
 654,326,341  & .216769   & .0630532 \\
\hline
std. dev. & .0001026 & .000069 \\
\end{tabular}
\caption{Estimated 
Bures/SD probabilities, based on Monte Carlo simulations, 
 that a three-qubit EW-state is: (1)
biseparable with respect to the partition $1|23$; and (2) triseparable. 
For each of the five subsamples, one billion points were initially generated.}
\end{table}
So, the Bures/SD probabilities of biseparability and triseparability are
considerably 
{\it greater} in the three-qubit case than in the three-qutrit one. This
is as would be expected on the basis of past analyses \cite{paul1}.
\subsection{Supplementary Analyses}
Among the constraints that must be satisfied for an EW-state to be 
triseparable are \cite[p. 6]{egg},
\begin{equation}
0 \leq r_{-} \leq {1 \over 6}; \quad {1 \over 4} (1-2 r_{-}) \leq r_{+} \leq 1 -5 r_{-}.
\end{equation} 
If we integrate the normalized form of (\ref{e3}) over the 
full domain of parameters (\ref{e4}), except for imposing these constraints,
we obtain an exact {\it upper} bound on the SD/Bures probability that a
qutrit-qutrit-qutrit EW-state is triseparable of
\begin{equation}
{-70 -325 \sqrt{2} \csc^{-1}{\sqrt{3}} +224 \sqrt{5} 
\sin^{-1}{\sqrt{5 \over 6}} 
\over 400 \pi} \approx .177661.
\end{equation}
If we act similarly, but now set $r_{-} = 0$ and 
use the normalized form of (\ref{e2}),
we obtain
an exact upper bound of $27/64 = .421875
$ on the SD/Bures probability that
a qubit-qubit-qubit EW-state is triseparable.

We can also get an apparently
 quite weak exact upper bound on the SD/Bures probability
that a three-qutrit EW-state is biseparable, by simply imposing the
single constraint $0 \leq r_{-} \leq 1/3$ \cite[p. 9]{egg}.
This gives
\begin{equation}
{26 \sqrt{2} + 54 \csc^{-1}{\sqrt{3}} \over 27 \pi} \approx .825312
\end{equation}
(while our Monte Carlo simulations [Table I] indicate that the actual
Bures/SD probability is more on the order of .069).
Though, somewhat to our disappointment, we have not been able to obtain
{\it exact} Bures probabilities of separability for the EW-states,
T. Eggeling has indicated (App. II) that if one restricts one's consideration
to the permutation-invariant EW states (which are commutative), then the
required integrations can be performed analytically.

As an additional investigative probe into the SD/Bures geometry of
the EW-tripartite states, having $U \otimes U \otimes U$ 
symmetry, we have estimated --- first, for the three-qubit
case --- the ratio ($\approx .34398$) 
of the SD area of the boundary of the triseparable
states to the SD area of the boundary of the biseparable states.
To obtain this estimate, we extracted the $3 \times 3$ submatrix of the
$5 \times 5$ metric tensor ((\ref{gfirst})-(\ref{glast})) 
corresponding to the variables $r_{1},r_{2},r_{3}$, 
and then took the square root ($h$)
 of its determinant (having, of course, set
$r_{-}=0$, as is demanded in the $d=2$ case). 
Then, we
randomly generated values of $r_{1},r_{2},r_{3}$ and computed for each of
the various defining inequalities, in turn, that
value(s) of $r_{+}$ which would saturate each.
We then evaluated $h$ if and only if 
that set of $r_{1},r_{2},r_{3},r_{+}$ also 
satisfied 
(without necessarily saturating) all the remaining 
required inequalities. 
The estimate of .34398
is based on some 21,000,000 such points. 
(For the first 10,000,000 points, the estimate was .342709.) 
We have not explicitly ruled out the
possibility that the SD areas themselves 
of the boundaries of the triseparable and
biseparable states in question are unbounded in value, as preliminary
analyses appear to indicate is the case for the area of the EW-states
in general.)

Similarly, for all the higher-dimensional three-qudit cases
($d>2$), where we now 
extract the $4 \times 4$
submatrix corresponding to $r_{+},r_{1},r_{2},r_{3}$ and solve for the values
of $r_{-}$ that saturate the various constraints, in turn, 
we obtained (on the
basis of 18,000,000  points) an estimate of .0949602 
for the ratio of the SD
area of the boundary of triseparable states to the SD area of the boundary
of biseparable states, and an estimate of .0000105263  for the ratio of the
SD area of the boundary of triseparable states to the SD area of the
boundary of states having positive partial transposes (with respect to 
the first tensor factor). (For the first 9,000,000 points, the estimates were,
respectively, .0959525 and .000010494.)
\section{Concluding remarks}
Let us indicate a number of  papers of Batle 
{\it et al} \cite{batle1,batle2,batle3}. 
In particular, 
these authors have 
investigated  --- following the review paper of Terhal \cite{terhal} --- the 
question of estimating the probabilities of (bi-)separability,
 using {\it various} separability criteria \cite{batle3}.
 They have {\it not}
employed the Bures measure in their various studies, as we have done here
and elsewhere \cite{paul1,paul2,paul3,paul4},
 but rather the (ZHSL) one utilized by \.Zyczkowski, Horodecki, Sanpera
and Lewenstein in their pioneering paper \cite{zycz1}. However, we have observed in \cite{lmp} 
that  the ZHSL 
 measure is not, in fact, proportional to the volume element of any monotone
metric, and, additionally, is
based on an {\it over}-parametrization of the $n^2-1$-dimensional convex set of $n \times n$ density matrices. So, one might argue, that the ZHSL measure 
serves best as a heuristic
device, easily computable, but lacking a 
suitably rigorous theoretical rationale.

Ac\'in {\it et al} \cite{acin} 
have introduced a classification of mixed {\it three-qubit}
states, in which they define the classes of separable, biseparable, $Q$ and
Greenberger-Horne-Zeilinger states. Among other things, they 
conclude in contrast to the
pure $W$-type states, that the mixed $W$ class is not of
measure zero since it is contained in a ball of finite radius
 (cf. \cite{braun2,szarek}).
Following the classification in \cite{acin}, we have obtained in
 \cite{paul4} --- using quasi-Monte Carlo 
methods --- estimates of lower bounds on the Bures/SD probabilities of being a GHZ state 
and of being a $W$ state.

Clifton and Halverson \cite{clifton} gave an elementary proof that the
set of separable density matrices for any bipartite quantum system in
which either part is {\it infinite}-dimensional is trace-norm {\it nowhere}
dense in the set of all density operators (cf. \cite{gauss}).
 They assert that this result
complements investigations \cite{braun2} concluding that if both parts
are finite-dimensional, then there is a 
(norm-dense) separable neighborhood of the
maximally mixed state.
However, in the case of $4 \times 4$ density matrices, if one adopts
the {\it maximal} monotone metric, then it appears that the ratio of
the volume of separable states to that of separable and nonseparable
states is zero \cite{paul4}.
\section{Appendix I of T. Eggeling}
\epsffile{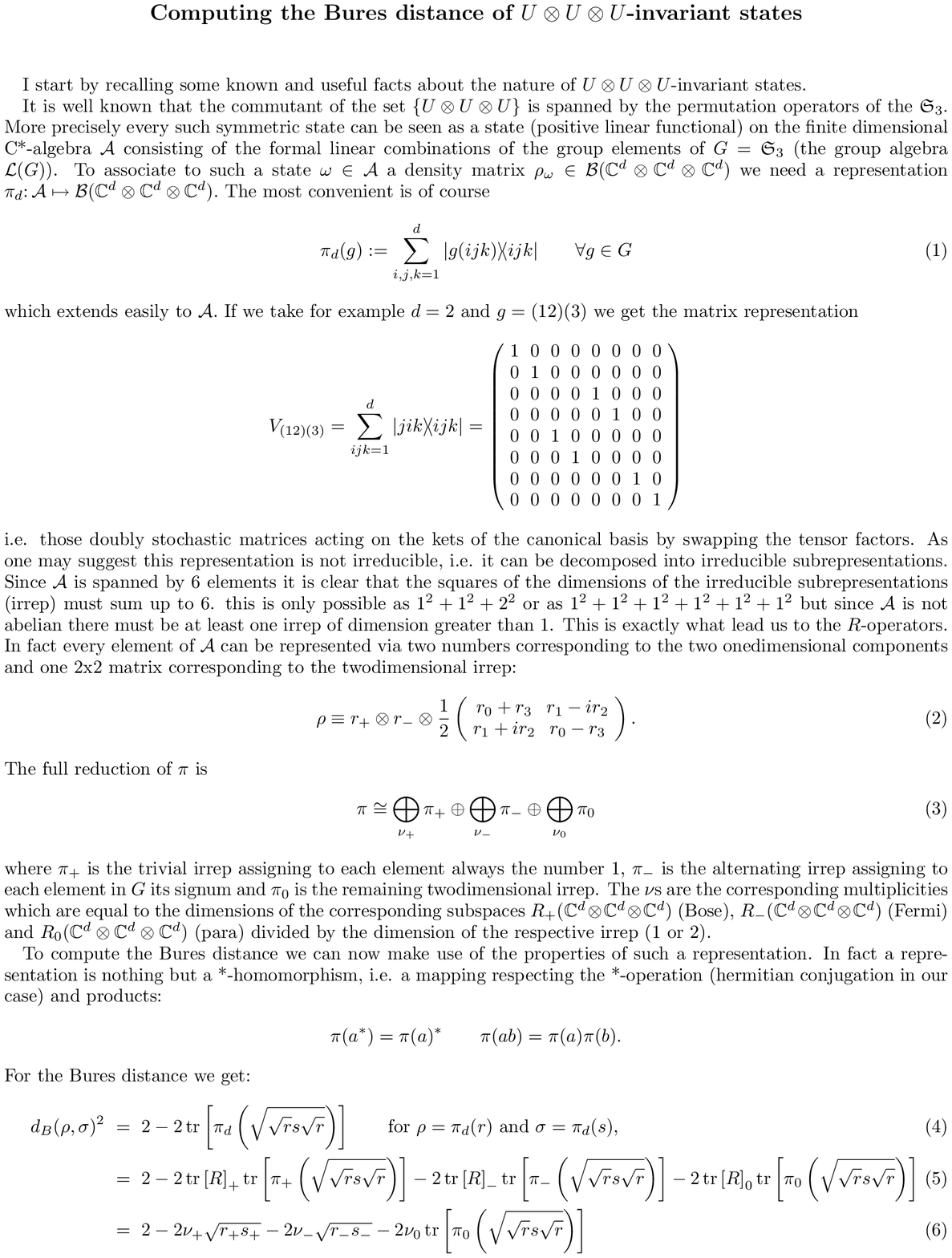}
\epsffile{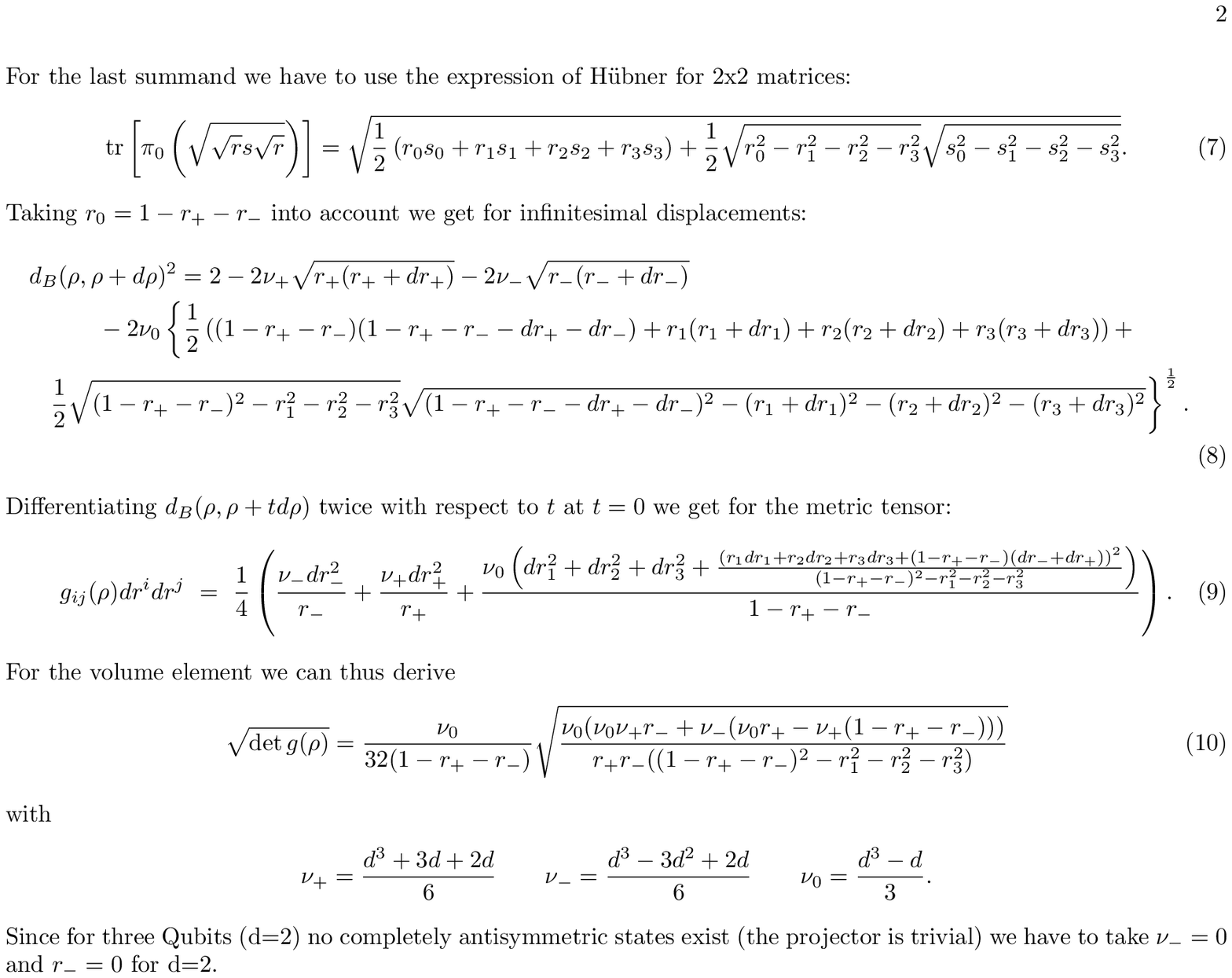}
\section{Appendix II of T. Eggeling}
For commutative subsets of states (=abelian algebras) one can write down 
the Bures distance directly. Therefore, taking the subset of 
permutation invariant tripartite Werner states 
(which is commutative) we can compute the volume element and integrate over 
the triangles you see in the picture of my old PRA paper. 
This integration can be done analytically and leads to: 
\begin{equation}
p_{sep,trip,perm} =
 {\pi \over 40} (-16 + 6 \sqrt{6} 
+ 5 \log{{3 (6-\sqrt{6}) \over 6 + \sqrt{6}}} = 0.170502,
\end{equation}
\begin{equation}
p_{bisep,trip,perm} = {\pi \over 10} (1-5 \sqrt{5}+4 \sqrt{6}
-10 \log{(5+\sqrt{5}) (6-\sqrt{6}) \over (5-\sqrt{5}) (6+\sqrt{6})}) 
= 0.179607,
\end{equation}
ppt=biseparable for permutation invariant tripartite states, so there is nothing to calculate.
These numbers together with the Monte-Carlo simulations correspond to the intuition that triseparable and separable states are concentrated in the vicinity of the subset of permutation invariant states 
This is intuitive if one looks at the figures describing these sets of states.

\acknowledgments
I would like to express appreciation to the Kavli Institute for Theoretical
Physics for computational support in this research and to T. Eggeling
for his sustained interest and involvement.

\end{document}